\documentclass[]{revtex4}
\usepackage{hyperref}
 
\begin{document}

\title{ Generalized BRST Symmetry and Gaugeon Formalism for Perturbative Quantum Gravity:  Novel Observation }

\author{ Sudhaker Upadhyay}
 \email {  sudhakerupadhyay@gmail.com; 
 sudhaker@boson.bose.res.in}

\affiliation { S. N. Bose National Centre for Basic Sciences,\\
Block JD, Sector III, Salt Lake, Kolkata -700098, India. }
 
\begin{abstract}
In this paper the novel features of  Yokoyama  gaugeon formalism are stressed out for the theory of perturbative quantum gravity in Einstein curved spacetime. The  quantum gauge transformations
for the theory of 
perturbative gravity are demonstrated in the framework of gaugeon formalism. These quantum gauge transformations lead to renormalized gauge parameter. Further,   we
analyse the BRST symmetric gaugeon formalism which embeds more acceptable Kugo-Ojima subsidiary 
condition.  Further, the BRST symmetry is made finite and field-dependent. Remarkably,  
the Jacobian of path integral under finite and field-dependent BRST symmetry
amounts to the exact gaugeon action in the effective  theory of perturbative quantum gravity.
  \end{abstract}
\maketitle

\section{  Introduction}
The understanding  of gravity as a quantum theory is one of the great challenges of 
physics. In search of full quantum theory of gravity  some attempts are made   by incorporating 
some modern concepts, namely, string theory and loop quantum gravity.
However, in the mean time, it was realized that
these modern concepts also meet some enormous
conceptual and technical problems.
Although the  quantum field theoretic approach was considered originally, it became  useless that days
because of severe difficulties. Nowadays, it might be worthwhile to
reconsider the quantum field theoretical approach when the quantum field theory on curved spacetime is  well established.
The  covariant quantum theory of gravity in curved spacetime in a usual perturbative approach begins  with the Einstein-Hilbert theory and expands the full Riemannian metric   around a constant background.
In this approach, the diffeomorphism invariance of the theory gets translated into a gauge symmetry of the fluctuation  \cite{sw} and hence, the problem of formulating the corresponding quantum field theory  in the
Einstein curved spacetime is conceptually no more different than the usual gauge theories.
The study of quantum field theory in curved spacetime (particularly  in   de Sitter spacetime)  has significant role
in inflationary cosmologies \cite{haw, hg, adl, aa}. 
The recently observed data indicates that the
rate of expansion of universe is such that it may approach de Sitter spacetime asymptotically \cite{per}. Further, 
  the gauge invariant  perturbative quantum gravity in curved space
   has founded great attempts to unify gravity with the Maxwell theory \cite{ein}. 
 The
   gauge invariant gravity models  have their relevance in certain string theories \cite{ch,da,ah}.

On the other hand, in the standard quantization of the gauge theories the gauge invariance at the quantum level
gets converted into fermionic rigid
 BRST invariance \cite{ht}.
 Such BRST symmetry plays an   important role in the proof of renormalizability  and unitarity  of the  gauge theories \cite{ht}. 
 The generalized   BRST symmetry, the so-called FFBRST symmetry,
 has also been studied which has great implications on gauge theories in flat as well as in
 curved spacetime   \cite{sdj,sdj1,rb,susk,jog,sb1,smm,fs,sud1,rbs,sudhak, rs,sudha}. 
   Recently, the BRST symmetry for the perturbative quantum gravity in
the  curved spacetime has been analysed  \cite{faiza, upa}
 and also has been generalized in the FFBRST framework \cite{sudha}.
  
However,  in the  quantization of gauge theories, one does
not investigate the gauge transformation at the quantum level as the theory 
does not exhibit quantum
gauge freedom. A quantum theory is defined only after fixing a suitable gauge
which is parametrize by a gauge parameter. Such gauge condition breaks the local gauge invariance.
 Hayakawa and Yokoyama have shown   that
  a   shift in gauge parameter occurs through renormalization 
  which affects the gauge-fixing condition  \cite{haya}.
Besides this, Yokoyama's gaugeon formalism \cite{yo0,yok,yo1,yo2,yo3} provides the
wider framework for quantization of gauge theories.  Within the gaugeon formalism, theory
admits the quantum gauge transformation under which the shift in the
gauge parameter occurs naturally which gets identified with the renormalized gauge parameter 
\cite{yo0}. The  idea behind gaugeon formalism is to study the  quantum gauge freedom by extending
  the configuration space with the introduction of
set of extra fields (so-called gaugeon fields) in the effective Lagrangian density.
Since the gaugeon fields are not physical fields as they do not contribute in physical processes, one needs to remove them. Yokoyama first time  removed them by putting the extra subsidiary  condition  of Gupta-Bleuler type
which has certain limitations \cite{yo0}. 
Further  extension of configuration space is made by introducing 
ghosts corresponding to gaugeon fields \cite{ki,mk} and by doing so the Gupta-Bleuler type restriction gets converted into the  Kugo-Ojima type restriction \cite{kugo, kugo1}. 
The gaugeon formalism has been studied extensively in many contexts
\cite{ki, mk, mk1, naka, rko, miu, mir1, mir2, sudha1,sudha2}. We analyse the gaugeon formalism  in the FFBRST framework. 

In this paper we consider the 
diffeomorphism  invariant classical gravity theory  in the Einstein curved    spacetime and discuss the 
  BRST symmetry  of the  perturbative quantum gravity in covariant gauge. 
  Further, we extend the 
  configuration space by introducing two gaugeon fields which describe the quantum gauge freedom. 
  Such extended Lagrangian density, called the Yokoyama Lagrangian density, possesses quantum gauge transformation under which 
this remains \textit{form} invariant. The form invariance of the Yokoyama Lagrangian density
leads to a natural shift in gauge parameter which may be identified with the
renormalized gauge parameter.
 As the gaugeon fields do not contribute in
physical processes, we
put an extra restriction of the Gupta-Bluler type on gaugeon fields. But the Gupta-Bluler type condition is not valid for all cases. Hence  we further extend the action  by introducing 
ghost fields for each 
gaugeon fields to improve the limitations of Gupta-Bleuler condition. This extended action respects 
both the BRST  symmetry
and quantum gauge symmetry.
Now, we generalize  the BRST transformation 
by making the parameter finite and field-dependent.
Under such generalized  BRST  transformation the
  functional integral does not remain invariant and therefore we calculate the 
  field-dependent Jacobian 
for the path integral.
  We therefore found that under  generalized
BRST transformation, with a particular choice of the
finite field-dependent parameter, the Jacobian of path integral leads to BRST symmetric gaugeon extended action.
Therefore, we claim that the 
extended action within gaugeon formalism  
can be constructed simply by calculating Jacobian under
generalized BRST transformation.

We organize the paper in following manner.
In section II, we discuss the  perturbative quantum gravity in gaugeon formalism. In Section III, we 
analyse briefly the mechanism of generalized BRST transformation. 
The emergence of gaugeon action through Jacobian   is calculated explicitly in section IV. In the last section we draw the final remarks. 
 
\section{ The perturbative quantum gravity in gaugeon formalism}
In this section, we develop the theory of
perturbative gravity manifestly in the quantum  gauge invariant framework. For this purpose, we analyse the perturbative gravity in the Yokoyama gaugeon formalism
which possesses the quantum gauge transformation.
Let us begin with
  the Lagrangian density for the classical gravity  
\begin{equation}
{\cal L}_{inv}  =   \sqrt{ - \tilde g}  (R- 2\Lambda ), \label{kin}
\end{equation}
where $ \tilde g$ is the   determinant of the full metric $\tilde g_{ab}$, $R$  is the Ricci scalar curvature and $\Lambda$ is the cosmological constant. The units  are adopted here such that $16\pi G=1$.

Further, we decompose the full metric into its background and  perturbation parts  as follows
\begin{equation}
\tilde g_{ab} =g_{ab}+h_{ab},
\end{equation}
where $g_{ab}$ is the fixed background metric 
and $h_{ab}$ describes the small perturbations around fixed metric.
Here the background metric $g_{ab}$ is assumed to satisfy the
equation of motion of \ref{kin}, that is, the Einstein equation $R_{ab}=\Lambda g_{ab}$.
 This small perturbation $h_{ab}$  is considered as a field which has 
  to be quantized. 
 Now, we expand  the Lagrangian density (\ref{kin}) to second order in $h_{ab}$ and 
 thus obtain the  Lagrangian density for the linearized
gravity as 
\begin{eqnarray}
{\cal L}_{inv} =\sqrt{-g}\left[\frac{1}{2}\nabla^b h^{ac}\nabla_a h_{bc}-\frac{1}{4}\nabla_a h_{bc}\nabla^a h^{bc}+\frac{1}{4} (\nabla^a h-2\nabla^b h^a_{\ b})\nabla_a 
h+\frac{1}{2}\Lambda\left(h_{ab}h^{ab} -\frac{1}{2}h^2\right)\right],
\end{eqnarray}
with $h=h^a_{\ a}$.
This   linearized  Lagrangian density possesses the following gauge invariance (up to total divergence)  
\begin{eqnarray}
\delta_\rho  h_{ab}&=&\nabla_a \rho_b +\nabla_b \rho_a,
\end{eqnarray}
where $\nabla_a$ denotes the background covariant derivative and $\rho_a$ is a vector field.
  Analogous to
the usual gauge theory where the Lagrangian remains invariant under gauge transformations,
we treat the theory of perturbative gravity as a gauge theory 
which is invariant under  coordinate transformations.
Now, according to the standard  quantization procedure, the gauge invariance of the effective theory reflects the redundancy in the physical degrees 
 of freedom.  These unphysical degrees of 
freedom lead to constraints in the canonical quantization  \cite{ht} and 
divergences in the generating functional  under the path integral quantization. To resolve this problem one needs to break the gauge invariance  by  fixing a gauge. In this case
we choose the following covariant gauge 
  \begin{equation}
G[h]_a=(\nabla^b h_{ab} -k\nabla_a h) =0,\label{gauge}
\end{equation}
where $k\neq  1$ is a gauge parameter. Since, for $k=1$, one of the conjugate momenta corresponding to $h_{ab}$ 
vanishes, this leads to divergence in the partition function. 
In this view  $k$  is written sometimes in terms of an arbitrary finite constant $\beta$ as $(1+\beta)/ \beta $ \cite{hig}.
 Now,  the gauge condition (\ref{gauge}) is incorporated 
by adding a covariant gauge-fixing term in the Lagrangian density as
follows  \cite{hig}  
 \begin{eqnarray}
{\cal L}_{gf}= \sqrt{- g}\left[ b^a(\nabla^b h_{ab}-k \nabla_a h) +\frac{\alpha}{2} b_ab^a\right], \label{gfix} 
\end{eqnarray} 
where $\alpha$ is also a gauge parameter and $b^a$ is a Lagrange multiplier field. To compensate
the contribution of this gauge-fixing term within the functional integral we add the following   Faddeev-Popov ghost term   
in the resulting action 
  \begin{eqnarray}
{\cal L}_{gh}&=&   \sqrt{- g}\bar c^a \nabla^b [ \nabla_a c_b+ \nabla_b c_a- 
2kg_{ab}\nabla_c c^c ], \nonumber\\
  &=&\sqrt{ -g}\bar c^a M_{ab} c^b,
\end{eqnarray} 
where the Faddeev-Popov matrix operator  $M_{ab}$   has the following form:
\begin{eqnarray} 
M_{ab}=   \nabla_c \left[ \delta_b^c\nabla_a  + g_{ab}\nabla^c - 2k \delta_a^c\nabla_b \right].
 \end{eqnarray}
Now, the total effective Lagrangian for perturbative quantum gravity in covariant gauge  can be given as the sum of gauge invariant, gauge-fixing and ghost terms  as follows
\begin{equation}
 {\cal L}_{T} = {\cal L}_{inv} +{\cal L}_{gf}+{\cal L}_{gh}.  \label{com}
\end{equation}
This effective Lagrangian density is invariant under following nilpotent BRST ($s$) 
transformation:
\begin{eqnarray}
&&s  h_{ab} =   \nabla_a c_b +\nabla_b c_a,\ \ s c^a  = -c_b\nabla^b c^a, \ \ s  \bar c^a
= b^a, \ \  
 s  b^a =  0. \label{sym}
\end{eqnarray}
 Utilizing this BRST symmetry transformation we ensure that the gauge-fixing and ghost terms of the complete Lagrangian density are BRST-exact and can be expressed as
\begin{eqnarray}
{\cal L}_g &:=& {\cal L}_{gf} +{\cal L}_{gh},\nonumber\\
&=& s  \Psi,\label{g}
\end{eqnarray}
 where  $\Psi$ is the gauge-fixing fermion for the theory of perturbative quantum gravity
 with the following expression:
\begin{equation}
\Psi =  \sqrt{ -g}\left[\bar c ^a \left(\nabla^b h_{ab} -k\nabla_a h +\frac{\alpha}{2} b_a\right)\right].\label{gff}
\end{equation}
Here we note   that any physical quantity does not depend on the choice of  gauge-fixing fermion \cite{ht}. However, the invariance of perturbative quantum gravity under BRST symmetry plays a
 crucial role in constructing the physical states of the theory.
\subsection{Yokoyama Gaugeon formalism }
In this subsection, we review the Yokoyama gaugeon formalism  for
perturbative quantum gravity in the Einstein curved spacetime as discussed in \cite{sudha1}. For this purpose, we start by constructing
the Yokoyama Lagrangian density for perturbative quantum gravity
 as
\begin{eqnarray}
{\cal L}_{yk}& =& {\cal L}_{inv} +\sqrt{- g} b^a(\nabla^b h_{ab}-k \nabla_a h)  +\frac{\varepsilon}{2} \sqrt{- g}(
y^a_\star + \lambda b^a )^2 
+\sqrt{ -g}\bar c^a M_{ab} c^b \nonumber\\
&+&  \sqrt{- g} \nabla^b y_\star ^a [ \nabla_a y_b+ \nabla_b y_a- 
2kg_{ab}\nabla_c y^c ],\label{yk}
\end{eqnarray}
which is an extended version of  total effective Lagrangian density (\ref{com}) having two extra gaugeon fields $y^a$ and $y^a_\star$   satisfying 
Bose-Einstein statistics. $\varepsilon(=\pm)$ is a sign factor  
and the original gauge parameter $\alpha$ (given in (\ref{gfix})) corresponds to $\varepsilon\lambda^2$.

Now,  the quantum gauge transformations for the Lagrangian density (\ref{yk}), under which the gauge parameter gets shifted, are demonstrated as 
 \begin{eqnarray}
&& h_{ab} \rightarrow\hat h_{ab} =   h_{ab} -\tau( \nabla_a y_b +  \nabla_b y_a),\nonumber\\
&&y^a_\star\rightarrow \hat y^a_\star = y^a_\star - \tau b^a,\nonumber\\
&& y_a\rightarrow\hat y_a =y_a, 
\nonumber\\
&& b^a\rightarrow \hat b^a =b^a,\nonumber\\
&& \bar c^a\rightarrow \hat{\bar c}^a  =\bar c^a,\nonumber\\
&& c^a\rightarrow \hat c ^a =c^a,  
  \end{eqnarray}
where the infinitesimal parameter of transformation  $\tau$ is bosonic in nature. Under these 
transformations the Lagrangian given in 
(\ref{yk}) is ``form invariant", i.e. it transforms as 
\begin{eqnarray}
{\cal L}_{yk}(\hat \phi,\hat \lambda) ={\cal L}_{yk}(  \phi, \lambda),
\end{eqnarray}
where $\hat\phi$ denotes the quantum gauge transformed  collective field $\phi$ and the
shifted parameter $\hat \lambda$ is defined by 
\begin{equation}
\hat \lambda = \lambda +\tau.
\end{equation}
The form invariance implies that the quantum fields $\hat \phi$ and $ \phi$ satisfy the same 
equations of motion
with gauge parameters $\hat \lambda$ and $\lambda$ respectively. 

It is easy to check that the Yokoyama Lagrangian density (\ref{yk}) is also invariant under 
the following nilpotent BRST  transformation 
\begin{eqnarray}
s  h_{ab} &=&    \nabla_a c_b +\nabla_b c_a, \ s  \bar c^a
= b^a\nonumber\\ 
s c^a  &= & -c_b\nabla^b c^a,\ \ \  s  b^a   =0,\nonumber\\
 s y^a & =&  0,\ \  \ s y^a_\star =0.  \label{bechhi}
\end{eqnarray}
Excluding the BRST variation of Yokoyama fields, the above symmetry transformations are identical to those in \ref{sym}) . 

To remove the unphysical modes and define physical states, we impose  two
subsidiary conditions:
\begin{eqnarray}
&&Q_b|\mbox{phys}\rangle =0,\nonumber\\
&& y_\star^{a(+)}|\mbox{phys}\rangle =0,\label{kogo}
\end{eqnarray}
where $Q_b$ is Noether's conserved charge corresponding to the BRST symmetry (\ref{bechhi}).
However, the  second subsidiary condition (of the Gupta-Bleuler type) works only when (i) the   background spacetime is flat, which means the cosmological constant should be zero and (ii) 
 the field $y^a_\star$  
satisfies the following
free field equation:
\begin{equation}
\nabla_b\nabla^b y^a_\star =0,
\end{equation}
which can be established by exploiting equations of motion of $y^a$.  
If the free field equation does not hold, the decomposition of 
field $y^a_\star$ into the positive and negative frequency
parts is no more valid. In Eq. (\ref{kogo}),
the  first Kugo-Ojima type condition is used to remove  the unphysical gauge field
modes from the total Fock space \cite{ko}; however, the second   Gupta-Bleuler
type condition  is used  for
the unphysical gaugeon modes.

\subsection{BRST symmetric gaugeon formalism}
In this subsection,  we discuss the BRST symmetric  gaugeon formalism for perturbative quantum gravity. 
For this purpose, we first extend the effective Yokoyama Lagrangian density 
by introducing two Faddeev-Popov ghosts $K_\star^a$ and $K^a$   corresponding to the gaugeon 
fields as
follows:
\begin{eqnarray}
{\cal L}_{ykb}& =& {\cal L}_{inv} +\sqrt{- g} b^a(\nabla^b h_{ab}-k \nabla_a h)  +\frac{\varepsilon}{2}\sqrt{- g} (
y^a_\star + \lambda b^a )^2 
\nonumber\\
&+& \sqrt{ -g}\bar c^a M_{ab} c^b + \sqrt{- g} \nabla^b y_\star ^a ( \nabla_a y_b+ \nabla_b y_a- 
2kg_{ab}\nabla_c y^c )\nonumber\\ 
&+& \sqrt{ -g}  K_\star^a (\nabla^b\nabla_a K_b +\nabla_b\nabla^b K_a
-2k g_{ab}\nabla^b\nabla_c K^c), \label{ykb}
\end{eqnarray} 
This extended Lagrangian density is invariant  under the
following  infinitesimal  BRST transformation:
 \begin{eqnarray}
&&s  h_{ab} =    \nabla_a c_b +\nabla_b c_a , \nonumber\\ 
&&s c^a  = -c_b\nabla^b c^a,\    \  \ s  \bar c^a
= b^a,\nonumber\\
&&   
 s  b^a =  0,  \ s y^a =-K^a,   \ sy^a_\star =0,   \nonumber\\
&& s K^a_\star  =y^a_\star,\ \ s K^a =0,\label{brs}
\end{eqnarray}
where gaugeon fields form BRST quartet. 
It is easy to check the nilpotency (i.e. $s^2 =0$) of the above BRST transformation. 

Thereafter, we see that the  sum of  gauge-fixing and ghost parts of the above Lagrangian density is BRST-exact and 
therefore can be written in terms of the BRST variation of extended gauge-fixing fermion $\Psi$ as
\begin{eqnarray}
{\cal L}_{ykb}& =& {\cal L}_{inv}  + s \Psi, 
\end{eqnarray} 
where the extended gauge-fixing fermion has the following expression 
\begin{eqnarray}
\Psi & =& \sqrt{- g} \left[\bar c^a \left(\nabla^b h_{ab} -k\nabla_a h\right) +\frac{\varepsilon\lambda}{2} \bar c^a \left(y_{a\star} +\lambda b_a\right)\right.\nonumber\\
&-&\left. K_\star^a \left(
\nabla^b\nabla_a y_b +\nabla_b\nabla^b y_a
-2k g_{ab}\nabla^b\nabla_c y^c \right)+\frac{\varepsilon}{2} K_\star^a\left(y_{ \star a} +\lambda b_a\right)\right]. 
\end{eqnarray}
This extended gauge-fixing fermion depends on both graviton and gaugeon fields.
However, for vanishing  gaugeon fields  it identifies with the
original gauge-fixing fermion    (\ref{gff}). 

Now,  with the help of  Noether's charge ($Q$) corresponding to the above 
BRST symmetry, we define the Kugo-Ojima type physical subsidiary condition 
\begin{eqnarray}
 Q |\mbox{phys}\rangle =0.\label{ku}
\end{eqnarray}
This single subsidiary condition  removes  both the unphysical gaugeon and unphysical
graviton (gauge) modes from the physical subspace of states.
 For example, it can be seen from
expression (\ref{brs}) that the gaugeon fields $y, y_\star, K$ and $K_\star$ form
a BRST quartet  which  appear only as zero-normed states
in the physical subspace \cite{ko}. 
Therefore, the condition (\ref{ku})  works fine and does not have
any kind of limitations like the Gupta-Bleuler one. 

Further, the  extended Lagrangian density given in  (\ref{ykb}) 
remains form invariant under following   quantum gauge transformations
\begin{eqnarray}
&& h_{ab} \rightarrow\hat h_{ab} =   h_{ab} -\tau( \nabla_a y_b +  \nabla_b y_a ),
\nonumber\\
&&y^a_\star\rightarrow \hat y^a_\star = y^a_\star - \tau b^a,\nonumber\\
&& y_a\rightarrow\hat y_a =y_a, 
\nonumber\\
&&b^a\rightarrow \hat b^a =b^a,\nonumber\\
&& \bar c^a\rightarrow \hat{\bar c}^a  =\bar c^a,\nonumber\\
&& c^a\rightarrow \hat c ^a =c^a
+\tau K^a,\nonumber\\
&& K^a_\star \rightarrow  \hat K^a_\star = K^a_\star -\tau \bar c^a,\nonumber\\
&&   K^a \rightarrow  \hat K^a = K^a, \nonumber\\
&& \lambda \rightarrow \hat \lambda = \lambda +\tau.
 \end{eqnarray}
These transformations commute with the
BRST transformation given in  (\ref{brs}). Consequently, the BRST charge $Q$ also  remains
unchanged under the above quantum gauge transformations.
Therefore, the physical
space of states   annihilated by charge  remains intact with
these transformations. Hence, we conclude that
 the physical Hilbert space of the theory  
remains unchanged under these quantum gauge transformations.

\section{ The generalized BRST transformation}
In this section, we analyse the generalization of usual BRST transformation 
by making the infinitesimal BRST parameter finite and field-dependent.
Such a generalized BRST transformation is known as the finite-field-dependent BRST (FFBRST)
transformation \cite{sdj}.
The methodology of FFBRST transformation is as follows. We first define
 the usual BRST transformation
 characterized by  a Grassmann parameter $\delta\Lambda$, written compactly as
\begin{equation}
\delta_b \phi = {\cal R}[\phi] \delta \Lambda,\label{use}
\end{equation}
  where ${\cal R}[\phi]$ is the Slavnov variation of collective field $\phi$. However, the properties of this transformation do not depend on whether 
the parameter $\delta\Lambda$  is (i) finite or infinitesimal, (ii) field-dependent or not, as long 
as it is anticommuting and space-time independent. These observations give us liberty to 
  make the  infinitesimal parameter  $\delta\Lambda$ finite and field-dependent without
 affecting its Grassmannian nature.  To generalize the BRST transformation (\ref{use}) we 
start 
by making the  infinitesimal parameter field-dependent with the introduction of an arbitrary parameter $\kappa\ 
(0\leq \kappa\leq 1)$.
We allow the fields, $\phi(x,\kappa)$, to depend on  $\kappa$  in such a way that $\phi(x,\kappa =0)=\phi(x)$ and $\phi(x,\kappa 
=1)=\phi^\prime(x)$, the transformed field.

The usual infinitesimal field-dependent BRST transformation thus can be 
constructed generically as 
\begin{equation}
\frac{d\phi(x,\kappa)}{d\kappa} ={\cal R}[\phi(x)]  \Theta^\prime [\phi (x,\kappa ) ]
\label{diff}
\end{equation}
where the $\Theta^\prime [\phi (x,\kappa ) ]{d\kappa}$ is the infinitesimal but field-dependent parameter.
Now, the FFBRST transformation ($\delta_f $) is constructed by integrating the above  transformation from $\kappa =0$ to $\kappa= 1$, as follows
\begin{equation}
\phi^\prime\equiv \phi (x,\kappa =1)=\phi(x,\kappa=0)+{\cal R}[\phi(x)] \Theta[\phi(x) ],
\end{equation}
which can further be written as 
\begin{equation}
\delta_f \phi  = \phi' (x)-\phi(x) ={\cal R}[\phi(x)] \Theta[\phi(x) ],
\label{kdep}
\end{equation}
where 
\begin{equation}
\Theta[\phi(x)]=\int_0^1 d\kappa^\prime\Theta^\prime [\phi(x,\kappa^\prime)],
\end{equation}
 is the finite field-dependent parameter. 
Following the   procedure discussed above,  the FFBRST transformations for the
 perturbative quantum gravity  are constructed from the the infinitesimal BRST transformation (\ref{brs}) as
 \begin{eqnarray}
&&\delta_f  h_{ab} =    (\nabla_a c_b +\nabla_b c_a) \Theta[\phi(x)], \nonumber\\ 
&&\delta_f c^a  = -c_b\nabla^b c^a \Theta[\phi(x)], \nonumber\\
&& \delta_f  \bar c^a
= b^a \Theta[\phi(x)],\ \delta_f  b^a =  0,\nonumber\\
&&   \delta_f y^a =-K^a \Theta[\phi(x)],\  \ \delta_f y^a_\star =0,   \nonumber\\
&& \delta_f K^a_\star  =y^a_\star\ \Theta[\phi(x)],\ \ \delta_f K^a =0, 
\end{eqnarray}
where $ \Theta[\phi(x)]$ is an arbitrary finite field-dependent parameter.
Such FFBRST transformation with the finite field-dependent
 parameter is the symmetry  of the effective action. However, the 
path integral measure is not invariant under such transformation leading to 
field-dependent Jacobian \cite{sdj}.

The Jacobian   $J(\kappa )$  of the path integral measure $ {\cal D}\phi $ in the functional 
integral for such transformations is then evaluated for the  arbitrary finite field-dependent parameter  $\Theta[\phi(x)]$  as
\begin{eqnarray}
{\cal D}\phi^\prime &=&J(
\kappa) {\cal D}\phi(\kappa),
\end{eqnarray}
where the Jacobian can be replaced (within the functional integral) as
\begin{equation}
J(\kappa )\rightarrow \exp[iS_1[\phi(x,\kappa) ]],\label{s1}
\end{equation}
for some local functional $ S_1[\phi ]$, if and only if  the following condition is satisfied \cite{sdj}
\begin{eqnarray}
 \int {\cal{D}}\phi (x) \;  \left [ \frac{1}{J}\frac{dJ}{d\kappa}-i\frac
{dS_1[\phi (x,\kappa )]}{d\kappa}\right ]\exp{[i(S_{eff}+S_1)]}=0.\label{mcond}
\end{eqnarray}
 This condition is very 
crucial which preserves the consistency of transformed   functional integral
with original functional integral. 

Using the Taylor
expansion, the infinitesimal change in  $J(\kappa)$ is derived as follows (for explicit derivation see e.g. \cite{sdj})
\begin{equation}
\frac{1}{J(\kappa)}\frac{dJ(\kappa)}{d\kappa}=-\int d^4x\left [\pm {\cal R}[\phi(x)]\frac{
\partial\Theta^\prime [\phi (x,\kappa )]}{\partial\phi (x,\kappa )}\right],\label{jac}
\end{equation}
where $+$ sign is used for   bosonic fields and $-$ sign is used for fermionic fields.

Henceforth, we  observe that
under the FFBRST transformation with the field-dependent parameter $\Theta$, the functional 
integral transforms as 
\begin{eqnarray}
\int D\phi\ e^{i\int d^4x {\cal L}_T}\stackrel{FFBRST}{---\longrightarrow} \int J(\kappa) 
 D\phi\ e^{i\int d^4x {\cal L}_T} \equiv \int  
 D\phi\ e^{i\int d^4x {\cal L}_T+iS_1[\phi]},
\end{eqnarray} 
where ($ \int d^4x {\cal L}_T+ S_1[\phi]$) is an extended  effective action.
It means under FFBRST transformation  the original  effective action of linear gravity ($ \int d^4x {\cal L}_T$) gets transformed into an extended effective action.
However, to produce the  extra piece $S_1[\phi, \varphi]$ in the effective action having some extra
fields $\varphi$  through the Jacobian calculation, we first insert a well-defined path integral measure 
corresponding to the extra fields (i.e. $\int D\varphi$)
in the  functional integral by hand  before performing FFBRST transformation.
\section{Emergence of BRST symmetric Yokoyama gravity theory}
   In this section, we   show that for a particular choice of
   finite field-dependent BRST parameter the Jacobian of path integral measure
   leads to the gaugeon term   (within a functional integral) in the effective theory of the perturbative gravity  naturally.
   For this purpose, our specific choice of finite field-dependent parameter 
   is obtainable  from the following   infinitesimal field-dependent parameter   
 \begin{eqnarray}
 \Theta' [\phi,\varphi] &=&i\gamma\int d^4 y\sqrt{- g} \left[ \frac{\varepsilon\lambda}{2} \bar c^a (y_{a\star} +\lambda b_a) 
- K_\star^a \left( 
\nabla^b\nabla_a y_b +\nabla_b\nabla^b y_a
-2k g_{ab}\nabla^b\nabla_c y^c \right)\right.\nonumber\\
&-&\left.\frac{\varepsilon}{2}  K_\star^a \left(y_{ \star a} +\lambda b_a \right)\right], 
 \end{eqnarray}
 where  an arbitrary parameter $\gamma$ is considered to make it more general
 and $\varphi$ corresponds to the gaugeon fields ($y,y_\star, K, K_\star$) collectively.
 Furthermore,
 we calculate the infinitesimal change in Jacobian 
 for this particular choice of field-dependent parameter as follows
  \begin{eqnarray}
 \frac{1}{J(\kappa)}\frac{dJ(\kappa)}{d\kappa}&=&-i \gamma\int d^4 x\sqrt{- g}\left[ 
 -\frac{\varepsilon}{2}\lambda b_a(y^a_\star +\lambda b^a)+y^a_\star(
 \nabla^b\nabla_a y_b +\nabla_b\nabla^b y_a -2kg_{ab}\nabla^b\nabla_c y^c)\right.\nonumber\\
 &-&\left. K_\star^a ( \nabla^b\nabla_a K_b +\nabla_b\nabla^b K_a -2kg_{ab}\nabla^b\nabla_c K^c)-\frac{\varepsilon}{2}y^a_\star(y_{\star a} +\lambda b_a)\right],\nonumber\\
 &=& i \gamma\int d^4 x\sqrt{- g}\left[ 
  \frac{\varepsilon}{2}  (y^a_\star +\lambda b^a)^2 +\nabla^by^a_\star( 
\nabla_a y_b +\nabla_b y_a -2kg_{ab} \nabla_c y^c)\right.\nonumber\\
 &+&\left. K_\star^a ( \nabla^b\nabla_a K_b +\nabla_b\nabla^b K_a -2kg_{ab}\nabla^b\nabla_c K^c) \right],\label{j}
 \end{eqnarray}
 where we have utilized the formula (\ref{jac}).
 Now, we make an ansatz for the local functional $S_1$ which appears in the 
 exponential of Jacobian given in (\ref{s1}) as 
 \begin{eqnarray}
 S_1 [\phi(x,\kappa), \varphi(x,\kappa), \kappa]  &=&\int d^4 x \left[\xi_1 (\kappa) 
 (y^a_\star +\lambda b^a)^2+\xi_2 (\kappa)\nabla^by^a_\star( 
\nabla_a y_b +\nabla_b y_a -2kg_{ab} \nabla_c y^c)\right.\nonumber\\
 &+&\left.\xi_3(\kappa)  K_\star^a ( \nabla^b\nabla_a K_b +\nabla_b\nabla^b K_a -2kg_{ab}\nabla^b\nabla_c K^c) \right],
 \end{eqnarray}
 where $\xi_i, i=1,2,3$ are arbitrary $\kappa$-dependent constants
and satisfy initial boundary conditions 
 $\xi_i (\kappa=0)=0$. Now, exploiting relation (\ref{diff}),  the
 infinitesimal change in above functional
 with respect to $\kappa$ is given by 
 \begin{eqnarray}
 \frac{dS_1  }{d\kappa}&=&\int d^4 x \left[\frac{d\xi_1}{d\kappa} 
  (y^a_\star +\lambda b^a)^2+\frac{d\xi_2}{d\kappa} \nabla^by^a_\star( 
\nabla_a y_b +\nabla_b y_a -2kg_{ab} \nabla_c y^c)\right.\nonumber\\
 &+&\left.\frac{d\xi_3}{d\kappa}   K_\star^a ( \nabla^b\nabla_a K_b +\nabla_b\nabla^b K_a -2kg_{ab}\nabla^b\nabla_c K^c) \right.\nonumber\\
 &-&\left.\xi_2 (\kappa)\nabla^b y^a_\star( 
\nabla_a K_b +\nabla_b K_a -2kg_{ab} \nabla_c K^c)\Theta'
\right.\nonumber\\
 &+&\left.\xi_3 (\kappa) y^a_\star \Theta'( \nabla^b
\nabla_a K_b +\nabla^b\nabla_b K_a -2kg_{ab} \nabla^b\nabla_c K^c) \right].
\label{s}
 \end{eqnarray}
Further, the condition for numerical consistency of the generating functional
  given in ({\ref{mcond}) together with
  Eqs. ( \ref{j}) and (\ref{s})
  yields  
 \begin{eqnarray}
&& \int d^4 x \left[\left(\frac{d\xi_1}{d\kappa}-\frac{\varepsilon}{2}\gamma \sqrt{- g}\right)
  (y^a_\star +\lambda b^a)^2+\left(\frac{d\xi_2}{d\kappa}-\gamma \sqrt{- g}\right)\nabla^by^a_\star( 
\nabla_a y_b +\nabla_b y_a -2kg_{ab} \nabla_c y^c)\right.\nonumber\\
& &+ \left.\left(\frac{d\xi_3}{d\kappa}-\gamma\sqrt{- g}\right)   K_\star^a ( \nabla^b\nabla_a K_b +\nabla_b\nabla^b K_a -2kg_{ab}\nabla^b\nabla_c K^c) \right.\nonumber\\
 &&- \left.(\xi_2 -\xi_3)\nabla^b y^a_\star( 
\nabla_a K_b +\nabla_b K_a -2kg_{ab} \nabla_c K^c)\Theta'
  \right] =0.
 \end{eqnarray}
 By equating the coefficients of various terms present in the above expression   from LHS to RHS,
 we get following linear differential equations:
 \begin{eqnarray}
 \frac{d\xi_1}{d\kappa}  -\gamma\frac{\varepsilon}{2}\sqrt{- g} &=&0,\nonumber\\  
  \frac{d\xi_2}{d\kappa} -\gamma\sqrt{- g} &=&0,\nonumber\\
\frac{d\xi_3}{d\kappa}  -\gamma\sqrt{- g} &=&0.
\end{eqnarray}
However, the non-local ($\Theta'$-dependent) terms vanish leading to
the following constraints to the parameters:
\begin{equation}
   \xi_2 - \xi_3 =0.
\end{equation}
The solutions of linear differential equations
satisfying the initial boundary conditions ($\xi_i|_{\kappa =0} =0, i=1,2,3$) are 
 \begin{eqnarray}
 \xi_1 (\kappa) = \frac{\varepsilon}{2}\sqrt{- g}\kappa,\ \ \xi_2(\kappa) = \sqrt{- g}\kappa,\
  \ \xi_3(\kappa) =  \sqrt{- g} \kappa,  
 \end{eqnarray}
 where we have set the arbitrary constant parameter $\gamma = 1$. 
 With these solutions,  $S_1[\phi(x,\kappa), \kappa]$ at $\kappa =1$ (under FFBRST transformation) have the following form:
 \begin{eqnarray}
S_1 [\phi(x),  \varphi(x)]_{\kappa=1} &=&\int d^4 x\sqrt{- g} \left[ \frac{\varepsilon}{2} 
 (y^a_\star +\lambda b^a)^2+   \nabla^by^a_\star( 
\nabla_a y_b +\nabla_b y_a -2kg_{ab} \nabla_c y^c)\right.\nonumber\\
 &+&\left.    K_\star^a ( \nabla^b\nabla_a K_b +\nabla_b\nabla^b K_a -2kg_{ab}\nabla^b\nabla_c K^c) \right].
 \end{eqnarray}
 Since $S_1 [\phi(x),  \varphi(x)]_{\kappa=1}$ appears at the exponent  of
  Jacobian (\ref{s1}).
Hence, this $S_1 [\phi(x), \varphi(x)]_{\kappa=1}$ (within functional integral) accumulates to the
 effective action for perturbative gravity given in (\ref{com}) leading to 
 an extended action as follows:
\begin{eqnarray}
\int d^4x {\cal L}_T + S_1 [\phi(x,1),  \varphi(x, 1),  1] =\int d^4 x{\cal L}_{ykb},
\end{eqnarray}
which is exactly the BRST symmetric gaugeon action of perturbative gravity.
It means that the Jacobian of measure of functional integral under generalized
BRST transformation leads to complete BRST symmetric  action of perturbative gravity
written in gaugeon fields.
\section{  Conclusions}
We have discussed the gauge invariance and  BRST symmetry of perturbative quantum gravity in 
the Einstein curved spacetime,  particularly, in the covariant gauge condition. 
Further, we have perused the Yokoyama gaugeon formalism for the  theory of quantum gravity 
which extends the effective action by incorporating the quantum (gaugeon) fields.  
This extended  action respects the quantum gauge transformation under which 
the gauge parameter gets shifted which is claimed as the renormalized gauge parameter by
drawing the analogy with ordinary gauge theory. 
Therefore, such an observation may help in studying  the renormalizability 
of the quantum theory of gravity.  
Although gaugeon fields discuss the quantum gauge freedom, 
it does not contribute to physical processes. So    we have removed them by employing
 the Gupta-Bleuler type subsidiary condition. 
But, unlike the Kugo-Ojima type condition the  Gupta-Bleuler condition  has  certain limitation.
Furthermore, this limitations have resolved with the construction of
 BRST symmetric gaugeon formalism  in which we further extend
 the configuration space by employing  corresponding   ghost fields too. This
 extended action possesses an extended BRST symmetry where gaugeon fields form
 BRST quartet.  Within this framework, 
two Yokoyama's physical subsidiary conditions get replaced by a single Kugo-Ojima type 
condition. We have demonstrated the 
quantum gauge transformation for the BRST symmetric 
gaugeon effective action too which commutes
with the extended BRST symmetry. 
As a result,   the physical Hilbert space for perturbative quantum gravity 
remains 
unchanged  under the quantum gauge transformations. 

Further we have generalized the extended BRST symmetry of the perturbative gravity
by allowing the parameter to be finite and field-dependent.
Doing so, we have found that such transformation  changes the
Jacobian of functional integral non-trivially. 
However, for a suitable choice of
finite field-dependent parameter the Jacobian has led to the gaugeon mode in the perturbative theory of gravity. We have established the results 
by explicit calculations.
This signifies that generalized BRST symmetry with an appropriate
parameter will be helpful to describe the theory in gaugeon mode.
It will be interesting to  generalize the quantum gauge transformation in the same fashion as
the generalized BRST transformation which may lead to some new results.


\begin{thebibliography}{0}

\bibitem{sw} S. Weinberg, ``Gravitation and Cosmology", (John Wiley and Sons, New
York-1972).
 
\bibitem{haw} S.W. Hawking  and G. F. Ellis,   ``The Large-Scale Structure Of Space-Time" (Cambridge: Cambridge University Press, 1973).
\bibitem{hg} A. H. Guth,   Phys. Rev. D 23,  347 (1981). 
\bibitem{adl} A. D. Linde,   Phys. Lett. B 108,  389 (1982).
\bibitem{aa} A. Albrecht  and P. J.  Steinhardt Phys. Rev. Lett. 48, 1220  (1982).
\bibitem{per} S. Perlmutter, G. Aldering, G. Goldhaber, R. A. Knop, and P. Nugent,
et al, Astrophys. J 517, 565 (1999).






\bibitem{ein} A. Einstein, ``The Meaning of Relativity, fifth edition (Princeton University Press, 1956).
\bibitem{ch} A. H. Chamseddine, Int. J. Mod. Phys. A 16, 759 (2001).
\bibitem{da}  T. Damour, S. Deser and J. McCarthy, Phys. Rev. D 47, 1541 (1993).
\bibitem{ah} A. H. Chamseddine, Commun. Math. Phys. 218, 283 (2001).



 \bibitem{ht} M. Henneaux and C. Teitelboim,  Quantization of gauge
systems  (Princeton, USA: Univ. Press, 1992).

\bibitem{sdj} S. D. Joglekar and B. P. Mandal, Phys. Rev. D 51, 1919 (1995).
\bibitem{sdj1}  S. D. Joglekar and B. P. Mandal, Int. J. Mod. Phys. A 17, 1279 (2002).
\bibitem{rb} R. Banerjee and B. P. Mandal, Phys. Lett. B 27, 488 (2000).
 \bibitem{susk}   S. Upadhyay,   S. K. Rai and B. P. Mandal,  J. Math. Phys.  {52}, {022301} (2011).
 \bibitem{jog} S. D. Joglekar and A. Misra, Int. J. Mod. Phys. A 15, 1453 (2000).
\bibitem{sb1} S. Upadhyay and B. P. Mandal,   Mod. Phys. Lett.   {A 25}, {3347} (2010);
  EPL 93, 31001 (2011); AIP Conf. Proc. 1444, 213 (2012); Eur. Phys. J.  {C 72},  2065 
(2012); Annals of Physics {327}, 2885 (2012). 
\bibitem{smm} S. Upadhyay, M. K. Dwivedi and B. P. Mandal, Int. J. Mod. Phys. A 28, 1350033 (2013).
\bibitem{fs} M. Faizal, B. P. Mandal and S. Upadhyay, Phys. Lett. B 721, 159 (2013).
\bibitem{sud1} B. P. Mandal, S. K. Rai and S. Upadhyay, EPL 92, 21001 (2010).
\bibitem{rbs} R. Banerjee, B. Paul and S. Upadhyay,  Phys. Rev. D 88, 065
019 (2013).
\bibitem{sudhak} S. Upadhyay,     Phys. Lett. B 727,  293 (2013).
\bibitem{rs} R. Banerjee and S. Upadhyay,  arXiv:1310.1168 [hep-th].
\bibitem{sudha} S. Upadhyay,  Annals of Physics 340, 110  (2014). 
\bibitem{faiza} M. Faizal, Found. Phys. 41, 270  (2011).
\bibitem{upa} S. Upadhyay, Phys. Lett. B 723, 470  (2013).
 

\bibitem{haya} M. Hayakawa and K Yokoyama, Prog. Theor. Phys. 44,  533 (1970).

\bibitem{yo0} K. Yokoyama, Prog. Theor. Phys. 51, 1956 (1974).
\bibitem{yok}  K. Yokoyama,
Prog. Theor. Phys. 59, 1699 (1978); Prog. Theor. Phys. 60, 1167 (1978);  Phys. Lett. B 79, 79   (1978).
\bibitem{yo1} K. Yokoyama and R. Kubo, Prog. Theor. Phys. 52, 290 (1974).
\bibitem{yo2} K. Yokoyama, M. Takeda and M. Monda, Prog. Theor. Phys. 60, 927 (1978).
\bibitem{yo3} K. Yokoyama, M. Takeda and M. Monda, Prog. Theor. Phys. 64, 1412 (1980).
\bibitem{ki} K. Izawa, Prog. Theor. Phys. 88, 759 (1992).
\bibitem{mk} M. Koseki, M. Sato and R. Endo, Prog. Theor. Phys. 90, 1111 (1993).
\bibitem{kugo} T. Kugo and I. Ojima, Prog. Theor. Phys. Supplement No. 66,  1 (1979).
\bibitem{kugo1}
T. Kugo, I. Ojima, Nucl.Phys. B 144, 234 (1978).
\bibitem{sudha1} S. Upadhyay,  Eur. Phys. J. C 74, 2737 (2014). 
\bibitem{sudha2} S. Upadhyay,  EPL 105, 21001 (2014).
 \bibitem{mk1} M. Koseki, M. Sato and R. Endo,
  Bull. of Yamagata Univ., Nat. Sci. 14, 15 (1996).
\bibitem{naka} Y. Nakawaki,  Prog. Theor. Phys.  98,  5 (1997).
\bibitem{rko} R. Endo  and M. Koseki, Prog. Theor. Phys.  103,    3, (2000).
\bibitem{miu} H. Miura  and R  Endo, Prog. Theor. Phys. 117,  4, (2007).
\bibitem{mir1} M. Faizal, Commun. Theor. Phys. 57, 637 (2012).
\bibitem{mir2} M. Faizal, Mod. Phys. Lett. A 27, 1250147  (2012).


\bibitem{mirh} M. Faizal, A. Higuchi, Phys. Rev. D 78, 067502 (2008).

 
\bibitem{ko} T. Kugo and I. Ojima, Phys. Lett.  B 73,   459 (1978); Prog. Theor. Phys. 60, 
1869 (1978).


\bibitem{hig} A. Higuchi and S. S. Kouris, Class. Quant. Grav. 18, 4317 (2001).
\end{thebibliography}
\end{document}